\documentclass[manuscript,screen]{acmart}

\usepackage{listings}
\usepackage{xcolor}
\usepackage{amsmath}
\usepackage{tikz}
\usepackage{subfig}
\usepackage{mathtools}
\usepackage{amsthm}
\usepackage{multirow}
\usepackage{booktabs}
\usepackage{multirow} 
\usepackage{pifont}
\usepackage{tabularx}
\usepackage{sidecap}
\usepackage[capitalize,noabbrev]{cleveref}
\usepackage{xspace}
\usepackage{soul}
\usepackage{enumitem}

\usepackage{algorithmic}
\usepackage[vlined,ruled,commentsnumbered,linesnumbered]{algorithm2e}

\newcommand{\cmark}{\ding{51}}%
\newcommand{\xmark}{\ding{55}}%
\colorlet{listing-highlight}{yellow!50} 
\newcommand{\sys}{RoSeMary}
\newcommand*\circled[1]{\tikz[baseline=(char.base)]{
            \node[shape=circle,fill,inner sep=0.5pt] (char) {\textcolor{white}{#1}};}}

\newcommand{\Prv}{$\mathcal{P}$\xspace}
\newcommand{\Cir}{$\mathcal{C}$\xspace}
\newcommand{\Vrf}{$\mathcal{V}$\xspace}
\definecolor{failed}{gray}{0.7}
\definecolor{keywordcolor}{RGB}{0,0,180}
\definecolor{commentcolor}{RGB}{0,128,0}
\definecolor{stringcolor}{RGB}{163,21,21}
\definecolor{diffadd}{RGB}{180,230,180}
\definecolor{diffdel}{RGB}{255,230,230} 

\newcommand\revision[1]{\textcolor{black}{#1}}
\newcommand\rerevision[1]{\textcolor{black}{#1}}
\lstdefinestyle{pythonstyle}{
    language=Python,
    basicstyle=\fontsize{5.5}{7}\selectfont\ttfamily,
    keywordstyle=\color{keywordcolor}\bfseries,
    commentstyle=\color{commentcolor}\itshape,
    stringstyle=\color{stringcolor},
    showstringspaces=false,
    escapeinside={(*@}{@*)}, 
    breaklines=true,
    columns=fullflexible
}

\newcommand{\diffadd}[1]{%
  \begingroup
  \setlength{\fboxsep}{0.1pt}
  \colorbox{diffadd}{#1}%
  \endgroup
}

\AtBeginDocument{%
  }

\setcopyright{acmlicensed}
\copyrightyear{2026}
\acmYear{2026}
\acmDOI{XXXXXXX.XXXXXXX}
\acmConference[]{ACM Transactions on AI Security and Privacy}{
  2026}{}
\acmISBN{978-1-4503-XXXX-X/2018/06}

\begin{document}

\title{Robust Zero Knowledge Verifiable Watermarking of Code LLMs with ML/Crypto Co-Design}

\author{Ruisi Zhang}
\orcid{00000-0002-9554-8073}
\authornote{Both authors contributed equally to this research.}
\email{ruz032@ucsd.edu}
\author{Neusha Javidnia}
\orcid{0009-0007-4783-3180}
\authornotemark[1]
\email{njavidnia@ucsd.edu}
\affiliation{%
  \institution{University of California, San Diego}
  \city{La Jolla}
  \state{California}
  \country{USA}
}

\author{Nojan Sheybani}
\orcid{0000-0002-4329-0197}
\affiliation{%
  \institution{University of California, San Diego}
  \city{La Jolla}
  \state{California}
  \country{USA}}
\email{nsheybani@ucsd.edu}

\author{Noah Marosok}
\orcid{0009-0008-4995-9219}
\affiliation{%
  \institution{University of California, San Diego}
  \city{La Jolla}
  \state{California}
  \country{USA}}
\email{nmarosok@ucsd.edu}

\author{Ke Huang}
\orcid{0000-0002-1587-9877}
\affiliation{%
  \institution{San Diego State University}
  \city{San Diego}
  \state{California}
  \country{USA}}
\email{khuang@sdsu.edu}

\author{Farinaz Koushanfar}
\orcid{0000-0003-0798-3794}
\affiliation{%
  \institution{University of California, San Diego}
  \city{La Jolla}
  \state{California}
  \country{USA}}
\email{farinaz@ucsd.edu}

\renewcommand{\shortauthors}{Zhang et al.}

\begin{abstract}
Watermarking has emerged as a viable tool for protecting intellectual property (IP) in large language model (LLM)-generated code. A major challenge, however, is the low-entropy nature of code, which constrains the space for high-quality watermarking signatures satisfying the detectability-fidelity-robustness tri-objective. Legal verification requires disclosing the signature to third-party arbitrators, and alternative signatures need to be encoded for artifact reuse to avoid forgery and tampering, which further strains the already limited high-quality signatures and reduces code usability.
This paper introduces \sys{}, a novel framework that secures high-quality code LLM watermark through ML/Crypto co-design. The end-to-end framework: (i) preserves code functionality, (ii) enhances detectability and robustness by leveraging pre-trained CodeT5 for better feature extraction and adversarial training, and (iii) customizes zero-knowledge proofs for efficient and secure verification without disclosing signatures to third parties. Extensive evaluations show \sys{} achieves high detection accuracy, preserves code functionality, resists attacks, and enables efficient zero-knowledge verification.
\end{abstract}





\maketitle

\section{Introduction}\label{intro}
The AI-empowered code-generation LLMs, such as GitHub Copilot~\cite{copilot}, Qwen2.5-Coder~\cite{hui2024qwen2}, and Code Llama~\cite{roziere2023code}, generate high-quality code via user instructions. They assist software engineers with agile development and reduce production costs~\cite{tan2023copilot,cai2024f}. 
Developing such powerful models requires substantially more effort compared to natural languages, e.g., designing specialized tokenization modules~\cite{li2022competition,roziere2023code}, acquiring high-quality code training data~\cite{lu2021codexglue,puri2021codenet}, and enhancing the model's reasoning ability~\cite{zhu2024deepseek,hui2024qwen2}. Nevertheless, AI-generated code may be used for malicious purposes and raise ethical and legal concerns, e.g., plagiarizing code that violates academic integrity~\cite{cyphert2023generative,tan2024rethinking}, contributing vulnerable code to open-source repositories~\cite{panichella2024vulnerabilities,garg2024coupling}, deploying code in commercial software without acquiring licenses from LLM owners that infringes intellectual property~\cite{yang2024srcmarker,wang2025rtlmarker}, etc.

Watermarking is suggested as a key solution to regulate LLM-generated content by embedding invisible signatures onto the code~\cite{huo2024token,liu2024adaptive}. The watermark insertion is formulated as an optimization problem that requires a signature insertion to preserve artifact quality while being identifiable by watermark decoding modules.  
The existing watermarking solutions fall into two approaches: (i) inference-based watermarking and (ii) neural-based watermarking.  Inference-based watermarking~\cite{lee2023wrote,ning2024mcgmark} encodes watermarks by splitting vocabulary into green/red lists on high-entropy tokens and decoding the next token primarily from the green list. Such methods manipulate watermark insertion at the token level and do not consider syntactic validity, which can introduce compilation errors and corrupt the code functionality.
Neural-based watermarking SrcMarker~\cite{yang2024srcmarker} employs a neural network to encode watermarks in both syntactic and variable name feature spaces for more robust watermarks. Its shallow transformer-based network architecture trained from scratch, however, limits SrcMarker's code feature extraction ability to provide higher watermarking strength and results in reduced detectability.

\begin{figure}[t]
    \centering
    \includegraphics[width=0.75\columnwidth]{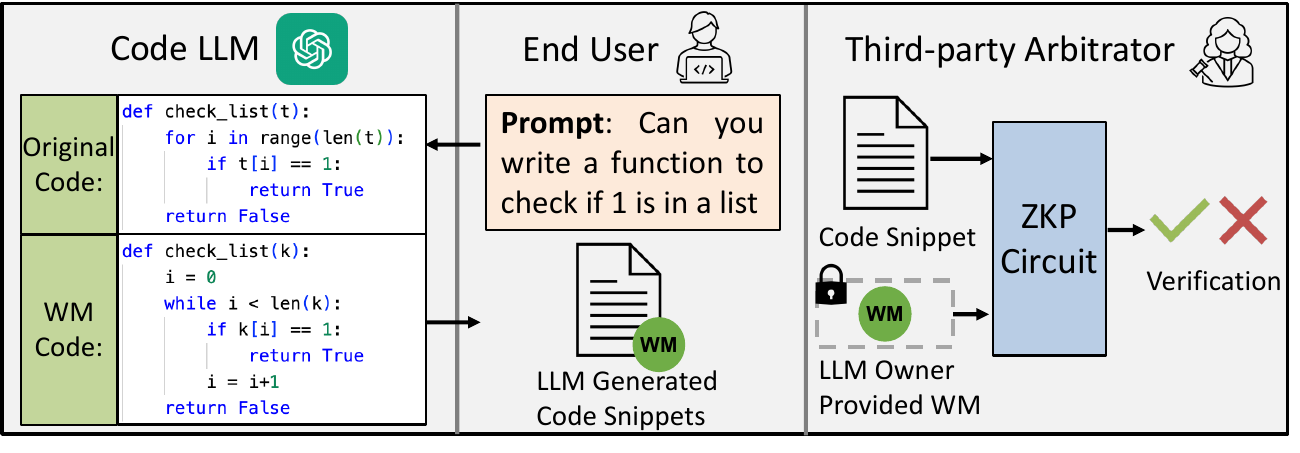}
    \caption{Overview of watermark insertion and extraction. The code LLM owner watermarks the code before distributing it to end users. The third-party arbitrator leverages zero-knowledge proofs to verify the ownership without requiring the owner to expose the encoded watermark, and provides legal credibility for potential IP misuse~\cite{liang2024watermarking}.}
    \label{fig:overview}
\end{figure}

To verify the ownership, existing schemes require LLM owners to disclose their signatures to third-party arbitrators for legal verification~\cite{kapusta2021protocol,liu2024survey}, which overlooks the reusability of watermarks. If owners want to use the same artifact, they must re-encode new watermarks to avoid signature forgery or exposure during the verification process. Such approach is practical for images or text, where the space of viable watermarking signatures is large.   However, for low-entropy code snippets, the limited watermarking space makes it difficult to generate alternative signatures for re-encoding without losing functionality or reducing watermark resilience.

\sys{} is a code LLM watermarking framework using ML/Crypto co-design that \revision{(i) jointly optimizes watermarking detectability, fidelity, and robustness with novel ML-based watermarking architecture design, and  (ii) enables reusing watermarked code with customized zero-knowledge proof protocols to keep signatures confidential during efficient verification.}
The Seq-to-Seq CodeT5~\cite{wang2021codet5}, pre-trained on millions of high-quality code snippets, is used as the watermark insertion backbone to extract better code features to fuse with watermarks and encourages watermark detectability. Then, it predicts a set of probabilities over the available syntactic transformations and variables to rename, which are applied to the original code to obtain the watermarked one.
A transformer decoder is used for watermark extraction. The watermark encoder and decoder are trained end-to-end to (i) preserve functionality by minimizing the code feature loss between original and watermarked code; and (ii) improve detectability and robustness by minimizing the message extraction loss between the encoded and extracted message from both the watermarked and the adversarially modified code.  As such, \sys{} co-optimizes the tri-objective for better watermarking performance.

As in Figure~\ref{fig:overview}, during \sys's deployment, the trained watermark encoder embeds the owner's signature into the LLM-generated code and distributes the watermarked one to users. Non-interactive zero-knowledge proofs (ZKPs) are used to verify ownership, allowing the computation-intensive proof generation to be performed once by the LLM owner while enabling an efficient and verifiable proof for subsequent watermark verifications by the arbitrator.
If a code snippet is suspected to be LLM-generated, users can submit the code to an arbitrator for legal inspection~\cite{kapusta2021protocol,liu2024survey} and request the LLM owner to input their signature to a ZKP circuit as a private input, \revision{which is not directly revealed to the arbitrator during verification}. \sys{} quantizes the watermark decoder to a smaller size and integrates both the watermark decoding and extraction rate computation into a unified computational graph for streamlined verification.
Hence, it enables efficient watermark source verification while keeping signatures private.

In brief, our contributions are summarized as follows:

\begin{itemize}
    \item \revision{Introduction of \sys{}, a code LLM watermarking framework that leverages ML/Crypto co-design to embed watermarking signatures into low-entropy code while supporting confidential ownership verification.}

    \item \revision{Leverage a pre-trained CodeT5 for better feature extraction and minimizes the code feature loss and message extraction loss between encoded and extracted signatures to encourage detectability of embedded signatures.}
        
\item \revision{Customizing and optimizing zero-knowledge proof circuit to enable efficient ownership verification without revealing encoded signatures.}

     \item Performing evaluations on extensive code benchmarks, demonstrating \sys{} (i) achieves 0.97 detection AUROC while preserving the code functionality and showing resilience against attacks, and (ii) enables efficient watermark verification without revealing underlying signatures within \rerevision{approximately 120 ms} using zero-knowledge proofs.
\end{itemize}



\section{Background and Related Work}\label{bg}
We first introduce the related work of LLM-generated code watermarking in Section~\ref{subsec:bg:llm_code_wm}, then review the background of zero-knowledge proofs (ZKPs) and optimizations for efficient verification in Section~\ref{subsec:bg:zkp}.

\subsection{Watermarking Large Language Model-generated Code}~\label{subsec:bg:llm_code_wm}
Compared to natural language, watermarking code needs to preserve both its semantics and functionality. Prior work can be methodologically categorized into two approaches~\cite{zhang2024remark}:  (i) inference-based watermarking~\cite{lee2023wrote,ning2024mcgmark}, and (ii) neural-based watermarking~\cite{yang2024srcmarker,wang2025rtlmarker}. 
The inference-based watermarking~\cite{lee2023wrote} encodes signatures during the LLM inference stage. It splits vocabulary into green/red lists only on high-entropy tokens and restricts LLM decoding to predict the next token from the green list. However, 
it manipulates watermarking insertion at the token level, which may violate syntactic validity and corrupt code functionality.    
Neural-based code watermarking approach~\cite{yang2024srcmarker} tries to maintain code functionality by encoding watermarks into the code using a neural network. It leverages a dual-channel network to embed watermarks on the code feature space and decodes a set of probabilities over potential syntactic transformations, as well as the variable to be renamed. Nevertheless, SrcMarker~\cite{yang2024srcmarker} employs a shallow transformer trained from scratch for watermark insertion/extraction, which limits code feature extractability and reduces watermark detectability. 

There is another line of work that employs rule-based methods like ACW~\cite{li2023protecting,li2024resilient} to watermark code. It maintains a transformation table containing the transformation ID and the rule to transform the code. For each code segment, rule-based watermarking applies available transformations on the original code to form the watermark and obtains the watermarked snippets. The watermarks can be vulnerable to removal attacks if the adversary has prior knowledge of the transformation table. Hence, we do not consider them in this paper.


Due to the low-entropy nature of code, the space for high-quality watermarks is constrained.
Once a watermark is disclosed for legal verification~\cite{kapusta2021protocol, liu2024survey}, re-embedding new signatures may compromise code functionality and resilience. 
Existing solutions primarily focus on improving watermark quality, but neglect secure verification mechanisms that protect the owner’s signature and avoid re-encoding watermarks that may harm code usability.


\subsection{Zero-knowledge Proofs for Machine Learning}~\label{subsec:bg:zkp}
Zero-knowledge Proofs (ZKPs) are a cryptographic primitive that allows a prover to prove knowledge of a secret value $w$ to a verifier. In a standard ZKP scheme, the prover $\mathcal{P}$ convinces a verifier $\mathcal{V}$ that $w$ is a valid private input such that $y=\mathcal{C}(x, w)$, in which $\mathcal{C}$ is an arbitrary computation and $x$ and $y$ are public inputs and outputs, respectively.
In general, ZKPs are extremely useful in computations where verification of outputs is costly (e.g. machine learning), as ZKPs allow users to verify a small proof rather than repeating the computation themselves \cite{xing2023zero}. In most ZK schemes, the majority of the computation lies in the setup and proving phases, as any computation $\mathcal{C}$ must be properly encoded in a way that ensures efficient processing during proof generation. In the context of ZK machine learning, for example, the layers, activation functions, and parameters, must all be represented as \textit{circuits}. This process, called \textit{arithmetization}, generally involves the conversion of the computations into arithmetic operations that can be efficiently performed over a finite field \cite{mouris2021zilch}. The setup and arithmetization phases of ZKP protocols are typically where the cryptographic elements are injected to ensure the privacy of $w$ \cite{hasan2019overview}.


Zero-knowledge proof generation can be performed in an interactive or non-interactive manner, depending on the application. One of the main drawbacks of interactive schemes is that they limit proofs to \textit{designated-verifier} settings, meaning proof generation, which is the most computationally heavy process in ZKP workflows, must be repeated for every new verifier. Non-interactive ZKPs allow for the \textit{publicly verifiable} setting, meaning that once a proof is generated attesting correct computation or valid data, it can be verified by any third party. Generally, non-interactive ZKP schemes can be represented with the three following algorithms: 

\begin{itemize}
    \item $(\mathcal{VK, PK})\xleftarrow[]{}$ Setup($\mathcal{C}$): A trusted third party, when trusted setup is needed, or \Vrf (with publicly verifiable randomness) runs a setup procedure to generate a prover key $\mathcal{PK}$ and verifier key $\mathcal{VK}$.
    \item $\pi \xleftarrow[]{}$ Prove($\mathcal{PK}$, \Cir, $x$, $y$, $w$): \Prv generates proof $\pi$ to convince \Vrf that $w$ is a valid witness. A malicious \Prv cannot generate a valid proof without knowledge of $w$. Alongside this, $\pi$ does not reveal anything about $w$.
    \item $1/0 \xleftarrow[]{}$ Verify($\mathcal{VK}$, \Cir, $x$, $y$, $\pi$): \Vrf accepts or rejects proof $\pi$. \Vrf cannot be convinced by an invalid proof due to soundness property of ZKPs \cite{tang2024zero}.
\end{itemize}

\revision{
Many prior works have explored the application of zero-knowledge proofs  in machine learning settings, including verifiable training~\cite{abbaszadeh2024zero,garg2023experimenting} and verifiable inference~\cite{qu2025zkgpt,ganescu2024trust}. These studies aim to improve the generalizability~\cite{lu2024efficient,chen2024zkml} and efficiency~\cite{sun2024zkllm,zhan2024validating} of ZKP-based protocols, with the goal of enabling a smooth user experience while reducing computational and communication overhead.
In confidential watermark verification, ZKROWN~\cite{sheybani2023zkrownn} applies ZKP to enable verifiable watermark extraction for watermarked DNNs. However, no prior work focuses on protecting the confidentiality of watermarks in LLM-generated code artifacts, which have a limited search space for high-quality watermarks. ZKP is particularly valuable for avoiding re-encoding signatures and thereby preserving the performance of watermarked code.}

\section{Problem Formulation}~\label{sec:threat}

In this section, we describe \sys's application scenario and motivation in Section~\ref{subsec:problem:scenario}, and present our threat model and potential attacks in Section~\ref{subsec:problem:threat}.

\subsection{Scenario and Motivation}~\label{subsec:problem:scenario}

Due to the \textbf{constrained high-quality watermark space} available in low-entropy code data, revealing signatures during legal verification and re-encode new ones to avoid forgery and tampering could otherwise degrade code usability.
As shown in Figure~\ref{fig:overview}, \sys{} encodes high-quality signatures onto the LLM-generated code before distributing it to users. A third-party legal arbitrator decides if the suspect code is from the owner utilizing a ZKP circuit. The owner's signature is a private input to the circuit, ensuring it remains confidential and known only to the owner.

\sys{} maintains the usability of the watermarked code throughout its life cycle with ML/Crypto co-design: (i) introducing a novel watermarking architecture that embeds signatures into code while preserving functionality, enabling high detectability and robustness against adversarial modifications; (ii) employing ZKPs for secure verification without revealing the watermark to third parties, thereby eliminating the need of re-encoding watermarks during code reuse. \revision{We formulate the required properties of an ideal watermarking system as follows:}






\begin{itemize}
    \item \textbf{Fidelity.} The watermarking process must not alter the semantics or functionality of the original code. The watermarked program should compile and execute correctly, producing identical outputs to the non-watermarked one. This ensures that watermarking remains unobtrusive and preserves the usability of the generated code.

    \item \textbf{Detectability.} The embedded watermark should be recoverable with high confidence from the watermark decoding module. Detectability ensures that ownership claims can be validated when required, enabling reliable attribution of LLM-generated code to its source.

    \item \textbf{Robustness.} The watermark extraction should remain reliable even when the code undergoes benign editing or malicious transformation, such as refactoring or variable renaming. A robust watermark resists removal or forgery attempts, preventing adversaries from erasing provenance information or falsely claiming ownership.

    \item \textbf{Reusability.} The legal ownership verification process should avoid exposing the embedded watermark. Otherwise, the disclosure would shrink the already limited watermark space in low-entropy code, thereby weakening security guarantees and harming the reusability of the protected artifact.

\end{itemize}

\subsection{Threat Model}~\label{subsec:problem:threat}

We consider the adversary to be a malicious end user aiming to remove or forge the watermark to deactivate the ownership proof. He/She can manipulate the watermarked code or discover the signature from the semi-honest third-party arbitrator. However, the adversary cannot control the \sys's watermark insertion or have access to \sys's network parameters or signatures. This setting is consistent with many prior arts~\cite{lee2023wrote,yang2024srcmarker}.

We consider the following attacks following previous work~\cite{lee2023wrote,yang2024srcmarker}: (i) Variable-rename Attack: the adversary randomly replaces variable names with semantically similar alternatives to remove the watermark while preserving code readability; (ii) Refactor Attack: the adversary uses a large language model to refactor the code in batches. This applies a combination of edits, including variable renaming and structural changes such as reformatting, control-flow restructuring, and adding wrappers (e.g., try/catch blocks); \revision{(iii) Re-Watermarking Attack: the adversary trains a malicious watermark system with a different set of hyperparameters. He/She wants to encode another set of new signatures while deactivating the proof of the original watermark}, and (iv) Forgery Attack: the adversary counterfeits another set of watermarks and falsely claims a non-LLM-generated code belongs to the owner.

Note that defenses against (i) adaptive attacks~\cite{lukas2023leveraging,diaa2024optimizing}, where adversaries iteratively modify content based on feedback from the arbitrator to remove watermarks, and (ii) spoofing attacks~\cite{sadasivan2023can,gloaguen2024discovering},  which alter LLM-generated content into undesired formats while preserving the watermark with the arbitrator's feedback, are orthogonal to \sys's objective to protect watermarks' confidentiality.
These attacks typically request verification tens to hundreds of times with similar content and can be mitigated via well-designed authentication protocols~\cite{kapusta2021protocol,pang2024attacking}, such as query rate limiting or anomaly detection. Thus, integrating these complementary defenses can further enhance the overall security of \sys{}.

\subsection{ZKP Security Guarantees}
\label{zkp_guarantees}
\revision{In this section, we provide an additional threat model formulation for ZKP. \sys{}'s verification is a non-interactive zero-knowledge proof of knowledge (NIZK) for the relation as follows:
    \[
    \mathcal{R}_{\mathbf{r}_e} = \bigl\{\,((e,\theta);\, M) \,:\, \mathrm{BER}
    \bigl(\mathbf{R}_e(e),\, M\bigr) \le \theta\,\bigr\},
    \]
The watermark extractor $\mathbf{R}_e(e)$ is hardwired into the circuit at setup time and the prover key $\mathcal{PK}$ and verification key $\mathcal{VK}$ commit to its parameters. Following Section~\ref{subsec:bg:zkp}, the underlying Halo2 based zkSNARK provides three properties: (i)~\emph{Completeness}: an honest LLM owner who knows a signature $M$ satisfying $\mathrm{BER}(\mathbf{R}_e(e), M) \le \theta$ always produces an accepting proof; (ii)~\emph{Knowledge soundness}: from any prover that convinces the verifier, an extractor recovers a witness $M\in\mathcal{R}_{\mathbf{R}_e}$, so a malicious actor cannot pass without actually possessing a qualifying signature; (iii)~\emph{Zero knowledge}: the proof $\pi$ is computationally indistinguishable from one produced by a simulator given only the public statement. As such, the verifier learns nothing about $M$ beyond the public output bit, making signature reusability possible.
    }
    
\noindent\textbf{\revision{Security Assumptions}} \revision{ \sys's ZKP end-to-end verification would be possible under the following assumptions: (1) an honest \texttt{Setup} that fixes the true $\mathbf{R}_e$ in $\mathcal{VK}$, (2)  a semi-honest arbitrator that follows the verification protocol, and (3) a fixed, agreed upon encoder $\mathbf{S}_e$ evaluated identically off-chain by both parties. All of these assumptions are consistent with the threat model in Section~\ref{subsec:problem:threat}.}

\revision{There are also several general proof-system level assumptions that \sys{} inherits from the standard KZG/Halo2 instantiation: (1) the hardness of the discrete logarithm problem in the pairing-friendly curve underlying the KZG polynomial commitment scheme, (2) an honestly generated universal SRS whose toxic waste has been destroyed, where we reuse a public perpetual ceremony rather than running a per-deployment setup, and  (3) Fiat-Shamir non-interactivity in the random oracle model.}


\noindent\textbf{\revision{Privacy Leakage}} 
\revision{By design, the protocol reveals exactly one bit of information about the secret: the public output $\mathsf{valid\_BER}\in\{0,1\}$, indicating whether the submitted code's extracted signature lies within Hamming distance $\theta$ of the owner's claims. The remaining public inputs $(\mathcal{S}(T,M)_{\mathsf{embed}},\,\theta)$ are also observable, but $\mathcal{S}(T,M)_{\mathsf{embed}}$ is a deterministic function of the already public watermarked code and therefore leaks nothing beyond it.  }

\revision{The verification key $\mathcal{VK}$ commits to $\mathbf{R}_e$ without revealing it, although reusing the same $\mathcal{VK}$ across proofs makes them linkable to the same decoder. \rerevision{In practice, this linkability has limited consequence for the owner: an observer learns only that two proofs were verified against the same committed decoder, i.e., that two disputes involve the same watermarking deployment. It reveals nothing about the signature $M$ and does not assist watermark removal or forgery. } A tight threshold $\theta$ constrains the secret signature to a small Hamming ball around $\mathbf{R}_e(e)$, bounding, but not eliminating, what the output bit implies about $M$.} 

\noindent\textbf{\rerevision{Scope of Security Guarantees}}
\rerevision{We distinguish three types of security threats and their protections. (i) \emph{Cryptographic Guarantees.} Signature confidentiality and knowledge soundness are guaranteed by the NIZK protocol under the assumptions stated above: an honest setup that commits the true decoder $\mathbf{R}_e$ in $\mathcal{VK}$, a semi-honest arbitrator, and the agreed-upon encoder evaluated identically off-chain. It ensures the code LLM owner's watermarks are kept confidential during one-time verification. (ii) \emph{Empirical robustness.} We perform extensive evaluations in Section \ref{subsec:robustness} to demonstrate \sys's resistance to variable renaming, LLM based refactoring, and re-watermarking attacks. (iii) \emph{Operational Defenses.} As discussed in Section \ref{subsec:problem:threat}, mitigating adaptive removal and spoofing attacks relies on access control around the verification service. Operational defenses, such as prover authentication and auditing, are orthogonal to \sys{} and can be incorporated to further strengthen its security.}

\section{Method}\label{method}

In this section, we first present the overall RoSeMary architecture in Section~\ref{subsec:method:architecture}; then detail the end-to-end training objectives for detectability, fidelity, and robustness in Section~\ref{subsec:method:training}; and finally introduce our zero-knowledge watermark verification for signature reusability in Section~\ref{subsec:method:verification}. 

\begin{figure*}[!ht]
    \centering
    \includegraphics[width=0.85\linewidth]{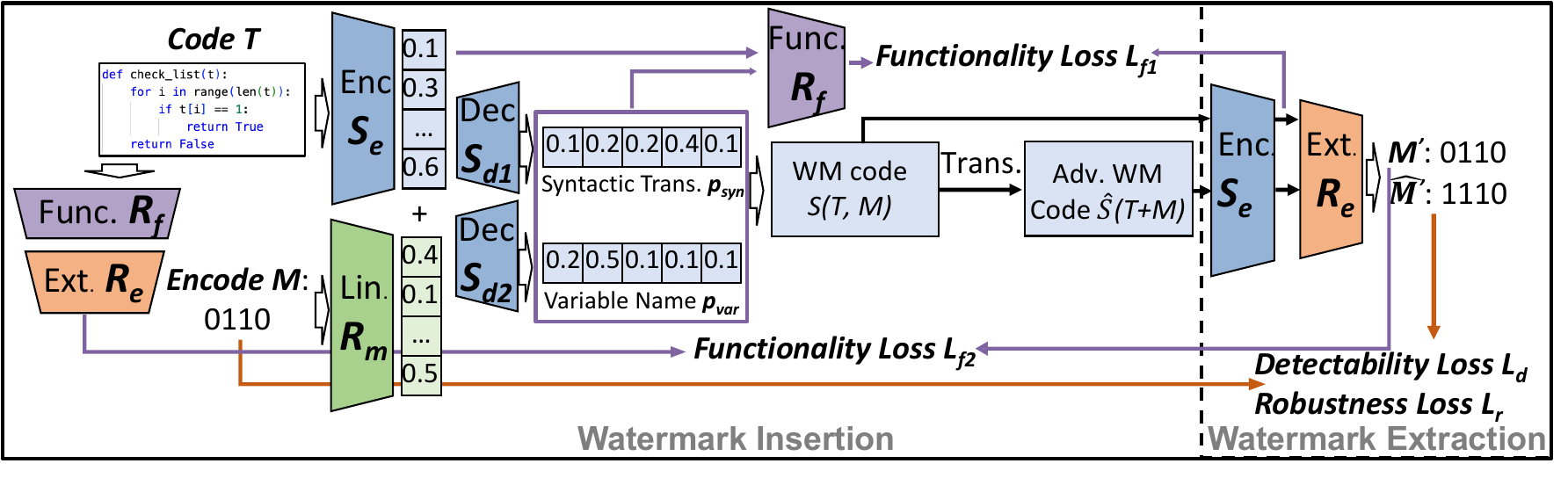}
    \caption{\sys{} watermarking overview. The \textbf{Watermark Insertion} takes the original code and watermark message as input and fuses their features by CodeT5's encoder $\mathbf{S}_e$. Two sets of decoders $\mathbf{S}_{d1}$ and $\mathbf{S}_{d2}$ predict the probability over the available syntactic transformations and the renamed variable over the vocabulary. Then, the \textbf{Watermark Extraction} module decodes watermarks from the syntactic-transformed and variable-renamed watermarked code $S(T, M)$, as well as its adversarial transformation $\hat{S}(T, M)$. The two parts are trained jointly to ensure (i) functionality-invariant by minimizing functionality loss $L_f$ and (ii) accurate and robust message decoding by minimizing detectability loss $L_d$ and robustness loss $L_r$.  }
    \label{fig:pipeline}
\end{figure*}


\subsection{\sys{} Architecture}~\label{subsec:method:architecture}
\sys{} consists of a watermark insertion and a watermark extraction module. As shown in Figure~\ref{fig:pipeline}, the watermark insertion backbone $\textbf{S}$ takes the watermark message $M$ and the code $T$ as input and generates (i) a probability over the syntactic transformations and (ii) variable name distribution over the vocabulary. Then, the code is watermarked by performing the transformations to get the watermarked code $S(T, M)$. Then, a watermark decoder extracts the message $M^\prime$ from the watermarked code $S(T, M)$. 


\textbf{Watermark Insertion} 
The watermark insertion employs the CodeT5~\cite{wang2021codet5}, pre-trained on millions of high-quality code files, as the backbone $\textbf{S}$ for watermark encoding. The encoder $\textbf{S}_e$ extracts the code feature and fuses with the message $M$'s feature extracted by $\textbf{R}_m$. The decoder $\mathbf{S}_{d1}$ and $\mathbf{S}_{d2}$ decodes two sets of probabilities over the syntactic transformations as $p_{syn}$ and variable token distributions $p_{var}$.  
Then, \sys{} obtains the watermarked code $S(T, M)$ by executing the predicted transformations from $argmax(p_{syn})$ and $argmax(p_{var})$. 

To approximate realistic post-processing edits without hard-coding a specific refactoring engine into the training loop, we generate adversarial variants by sampling alternative structure-preserving edit decisions from the model’s transformation distributions. As shown in Equation~\ref{eq:perturb}, the AST-level transformation probability distribution $p_{syn}$ and variable-renaming probability distribution $p_{var}$ are perturbed to obtain $\hat{p}_{syn}$ and $\hat{p}_{var}$. It encourages \sys{} to explore nearby but different edit sequences (e.g., alternative loop/condition forms, formatting-preserving AST rewrites, and different semantically valid renamings).
Applying these sampled edits yields an adversarially transformed code $\hat{S}(T, M)$ that preserves functionality but stresses watermark recovery, and we train the extractor to recover $M$ from both $S(T, M)$ and $\hat{S}(T, M)$. Note that this procedure does not aim to fully simulate arbitrary refactoring attacks, but provides a lightweight, differentiable way to cover a broad neighborhood of functionality preserving transformations that helps \sys{} to decode robust signatures.

 \begin{equation}
\label{eq:perturb}
\begin{array}{ll}
\hat{p}_{syn} &= p_{syn} + \epsilon_{syn} \\
\hat{p}_{var} &= p_{var} + \epsilon_{var}, \quad \epsilon_i \sim \mathcal{N}(0, \sigma_p^2)
 \end{array}
\end{equation}


\textbf{Watermark Extraction} The watermark extraction decodes messages from the watermarked code $S(T, M)$. The encoder $\textbf{S}_e$, with shared parameters and architecture as the watermark insertion, is used to extract code features. Then, a shallow decoder $\textbf{R}_e$ is used to recover message $M^\prime$. Similarly, the message $\hat{M}^\prime$ is decoded from the adversarial example $\hat{S}(T, M)$.

\textbf{Watermarking Strength} We measure the watermarking strength under the null hypothesis using a one-sided $z$-test on the total number of matching bits in Equation~\ref{eq:zscore}. The higher the z-score is, the more confident an owner can claim the code has been watermarked.
For an artifact with $F$ functions, \sys{} embeds watermark $M$ in each function.
$M$ is a binary sequence whose message generation is random, and follows binomial distributions. The probability for generating bit 0 is $p=0.5$, and bit 1 is $1-p = 0.5$. The mean of the message distribution is $\mu= F \times |M| \times p$, and the variance is $ \sigma^2=F \times |M| \times p \times(1-p)$. The watermark extraction accuracy is $q$, thus the expected matching bit is $q \times F \times |M|$.

\begin{equation}
\begin{aligned}\label{eq:zscore}
z=\frac{q \times F \times |M|-\mu}{\sigma}
\end{aligned}
\end{equation}

\subsection{\sys{} End-to-End Training}~\label{subsec:method:training}
We guide the training by minimizing the loss $L$ in Equation~\ref{eq:loss}. 
\sys{} is trained to meet three criteria: (i) Functionality-invariant: the functionality of watermarked code $S(T, M)$ remains the same as the input code $T$ as $L_f$; (ii) Detectability: the decoded message $M^\prime$ matches the encoded $M$ for successful detection as $L_d$; (iii) Robustness: the adversarial sample $\hat{S}(T, M)$'s decoded message $\hat{M}^\prime$ matches the encoded $M$ for robust detection as $L_r$.

\begin{equation}
\label{eq:loss}
L = w_f L_f + w_d L_d + w_r L_r
\end{equation}

\textbf{Functionality Loss} Performing transformations over $T$ results in non-differentiable watermarked $S(T, M)$. Inspired by SrcMarker~\cite{yang2024srcmarker}, we employ $\textbf{R}_f$ to approximate the watermark insertion/extraction. The $\textbf{R}_f$ encourages the code functionality feature to be close during training. This is achieved by minimizing the mean square error (MSE)~\cite{allen1971mean} $L_{f1}$ between original code feature $\textbf{R}_f (\mathbf{S}_e, p_{syn}, p_{var})$ and watermarked code feature $\textbf{S}_e (S(T, M))$ as in Equation~\ref{eq:functionality}. We also ensure the $\textbf{R}_f$'s approximation is correct for watermark extraction by minimizing the binary cross entropy (BCE) loss~\cite{ruby2020binary} $L_{f2}$ between $\textbf{R}_f(T)$ and predicted $M^\prime = \textbf{R}_e(\textbf{S}_e(S(T, M)))$.

\begin{equation}
\label{eq:functionality}
\begin{array}{rr}
L_f = MSE(\textbf{R}_f (\mathbf{S}_e, p_{syn}, p_{var}), \textbf{S}_e (S(T, M))) + \\ BCE(\textbf{R}_e(\textbf{R}_f(T)), \textbf{R}_e(\textbf{S}_e(S(T, M))))
\end{array}
\end{equation}

\textbf{Detectability Loss} \sys{} minimize the BCE loss between $M$ and $M^\prime$ in Equation~\ref{eq:detect} to recover correct message in watermark extraction.

\begin{equation}
\label{eq:detect}
L_d = BCE(M, M^\prime)
\end{equation}

\textbf{Robustness Loss} To enable robust message recovery over adversarial transformations, the watermark extraction also decodes the malicious message $\hat{M}^\prime$ over $\hat{S}(T, M)$ and minimizes the BCE loss between $M$ and $\hat{M}^\prime$ in Equation~\ref{eq:robust}.

\begin{equation}
\label{eq:robust}
L_r = BCE(M, \hat{M}^\prime)
\end{equation}


\subsection{Secure Watermark Verification}~\label{subsec:method:verification}


\textbf{\revision{Zero Knowledge Watermark Verification}}
The core problem with watermark extraction is the requirement to reveal the watermarked information to prove ownership of the code. This presents a significant challenge because each time a signature is exposed, the data must undergo re-encoding watermarks to prevent adversaries from discovering, altering, or erasing the exposed signature. Utilizing zero-knowledge proofs (ZKPs) we can solve this problem. We present a unique watermark extraction scheme, built using non-interactive ZKPs, that can efficiently prove that code has been generated using a proprietary LLM, without revealing what the original watermark was. Our solution generates verifiable proofs, such that one proof can be generated and universally verified to prove that a code snippet was generated from a proprietary LLM. 

We utilize Halo2-based zk-SNARKs \cite{halo2_repo}, a class of non-interactive ZKPs that offer high scalability and fast verification time. Our proposed system benefits from the fact that proof generation only has to be done once, and, as proof generation is the slowest aspect of Halo2-based zk-SNARKs, we do not need to view this as a bottleneck. Due to the computational overhead of ZKPs, our approach includes non-interactive ZKP-specific optimizations, such as custom quantization, to ensure that the operation is runtime and memory-efficient. 

To verify copyrights, the semi-honest legal arbitrator starts by mapping the watermarked code $\mathcal{S}(T, M)$ into its embedding space using the fixed feature extractor $\mathbf{S}_e$. This is done by running a feed-forward process on $\mathbf{S}_e$ with input $\mathcal{S}(T, M)$, which results in a feature vector $\mathcal{S}(T, M)_\textsf{embed}$ that represents the watermarked code in the correct embedding space.  A high-level approach towards our zero-knowledge watermark extraction scheme can be seen in Algorithm \ref{alg:extract}. The public inputs from the verifier (legal arbitrator) to the verification include $\mathcal{S}(T, M)_\textsf{embed}$ and a target bit error rate (BER) $\theta$. These public inputs do not reveal any sensitive information about the proprietary LLM or watermarking scheme.  Besides, the legal arbitrator provides a public digest of the submitted code $h(S(T,M))$, where the prover (LLM owner) checks $h(S(T,M))$ against the submitted code and re-computes $\mathbf{S}_e(S(T, M))$ off-chain to confirm it equals the public $\mathcal{S}(T, M)_\textsf{embed}$.

Then, the LLM owner begins the ZKP generation process.  The circuit is instantiated with a fixed shallow decoder $\bf{R}_e$ at the setup time such that the prover cannot modify $\bf{R}_e$ without invalidating the proof. The original signature $M$ is taken in as private inputs. With all computation represented as a zero-knowledge circuit, the trained watermark extraction module $\mathbf{R}_e$ decodes the signature $M^\prime$ from the watermarked code by performing our custom $zkFeedForward$ function on $\bf{R}_e$, with the input set to $\mathcal{S}(T, M)_\textsf{embed}$. This results in an extracted signature $M^\prime$ of the same length as $M$. Within the same zero-knowledge circuit, the BER between the extracted signature $M^\prime$ and the original signature $M$ that the LLM owner provides is calculated. This is done with our provided custom $zkBER$ function, which returns $1$ if the BER between $M^\prime$ and $M$ is less than $\theta$, or else it returns $0$. The resulting proof $\pi$ will only be valid if the extracted signature $M^\prime$ has a low enough BER compared to the original signature $M$. This proof $\pi$ can be sent to any verifier \Vrf to prove that the inspected watermarked code was a result of the model owner's proprietary LLM.

\begin{algorithm}[h]
\small
\caption{\revision{Non-Interactive ZK Watermark Verification. }}
\label{alg:extract}
\begin{algorithmic}[1]
    \STATE {\bfseries Public statement:} $x = (e,\,\theta)$, with $e = \mathcal{S}(T,M)_{\mathsf{embed}}$.
    \STATE {\bfseries Private witness:} $w = M$.
    \STATE {\bfseries Relation:}
    \[
    \mathcal{R}_{\mathbf{R}_e} \;=\; \bigl\{\,(x;\,w) \,:\, \mathrm{BER}\bigl(\mathbf{R}_e(e),\, M\bigr) \le \theta \,\bigr\},
    \]
    \STATE\hspace{1em} encoded by a fixed Halo2 circuit $C_{\mathbf{R}_e}$ in which the trained decoder $\mathbf{R}_e$ is hard-wired.

    \vspace{2pt}
    \STATE {\bfseries Setup}$(1^\lambda,\, C_{\mathbf{R}_e}) \to (\mathcal{PK},\,\mathcal{VK})$: \hfill // run once by a trusted setup
    \STATE \hspace{1em} arithmetize $C_{\mathbf{R}_e}$; emit proving/verification keys ($\mathcal{VK}$ commits to $\mathbf{R}_e$).

    \vspace{2pt}
    \STATE {\bfseries Prove}$(\mathcal{PK},\, x,\, w) \to \pi$: \hfill // run by the LLM owner
    \STATE \hspace{1em} {\bfseries return} $\pi \gets \mathsf{SNARK.Prove}(\mathcal{PK},\,x,\,w)$.

    \vspace{2pt}
    \STATE {\bfseries Verify}$(\mathcal{VK},\, x,\, \pi) \to \{0,1\}$: \hfill // run by the arbitrator
    \STATE \hspace{1em} {\bfseries return} $\mathsf{SNARK.Verify}(\mathcal{VK},\,x,\,\pi)$.
\end{algorithmic}
\end{algorithm}

\textbf{\revision{Efficiency Optimizations}}
A majority of the computational burden lies in the $zkFeedForward$, which requires custom optimization of the shallow  decoder to ensure efficient operation when translated to ZK computation. Specifically, $\bf{R}_e$ is made up of batch normalization, fully-connected, ReLU, and dropout layers. To implement and run $zkFeedForward$, we use a customized version of the EZKL Rust package \cite{ezkl}. EZKL accepts a computational graph as input, allowing us to optimize our computation before converting it to the correct input format. We provide four custom optimizations and capabilities to ensure efficient proof generation, while maintaining small proof size and fast verification:

\noindent\circled{1} We lower the memory requirement that is necessary for non-linear layers by adding support for polynomial approximations, which is an important technique in privacy-preserving applications. We approximate ReLU using $\sigma(x)=x^2+x$, which has been shown to closely replicate the ReLU \cite{ali2020polynomial}.

\noindent\circled{2} We quantize parameters into Bfloat16, a 16-bit floating point format that reduces the memory requirement for proof generation \cite{burgess2019bfloat16}, while keeping watermark extraction accuracy.

\noindent\circled{3} We add support for a highly efficient, zero-knowledge bit error rate calculation circuit based on the Halo2 proof system to represent $zkBER$. This is done using the bitwise AND operator to calculate the number of bits that differ between the $M^\prime$ and $M$.

\noindent\circled{4} We add a composability layer that allows for efficient combination of Halo2-based and EZKL circuits (e.g. $zkFeedForward$ and $zkBER$) for representation in a single computational graph.

Using these optimizations, we are able to build an efficient ZK watermark extraction and verification scheme with small proofs and fast verification that cleanly integrates into \sys{}'s end-to-end workflow.


\textbf{\revision{Circuit Design.}}
\revision{
The relation $\mathcal{R}_{\mathbf{R}_e}$ is encoded by a single arithmetic circuit $C_{\mathbf{R}_e}$ in which the trained shallow decoder $\mathbf{R}_e$ is fixed at setup time. Each $\mathrm{Linear}$ layer in $\mathbf{R}_e$ compiles to a fixed-coefficient inner product gadget over the field with its parameters baked into the circuit and committed by $\mathcal{VK}$. The activation $\sigma$ is the polynomial approximation $x^2 + x$ (optimization~\circled{1}) and therefore arithmetizes to pure multiply and add gates, with no range check lookup table required. The $zkBER$ block consumes the feed forward output $M'$, binarizes it bit-by-bit, computes the hamming distance to the private $M$ via per-bit XOR followed by a popcount accumulator, and \emph{enforces} (instead of returning) the constraint $\mathrm{BER}\le\theta$ inside the circuit. Because nonlinearities cost only $|M|$ bitwise gates, the resulting circuit fits in $2^{18}$ rows for the deployed configuration.
}


\section{Experiment}\label{exp}


\subsection{Experiment Setup}
\label{subsec:exp:setup}
\paragraph{\textbf{Dataset}} We use HumanEval~\cite{chen2021evaluating}, MBXP~\cite{austin2021program} (MBPP for Python, MBCPP for C++, and MBJP for Java), and EvalPlus~\cite{evalplus} as the target benchmark to evaluate \sys's performance. All of the datasets have instruction prompts for code generation, human-written canonical solutions, and test cases for functionality evaluation. 


\paragraph{\textbf{Evaluation Metrics}} We assess the frameworks' performance from: (i) \textbf{\textit{Detectability}}: detecting watermarked versus non-watermarked codes based on their respective z-scores, where ground-truth labels are set to 1 for watermarked and 0 for non-watermarked samples. AUROC is the area under the receiver operating characteristic curve, TPR is the true positive rates, and FPR is the false positive rates; (ii) \textbf{\textit{Fidelity}}: the percentage of code passed unit tests as Pass Rate.

\paragraph{\textbf{Baselines}}  We compare \sys{} with state-of-the-art natural language watermarking baselines: (i) \textbf{\textit{KGW}}~\cite{kirchenbauer2023reliability} is an inference-based watermarking scheme for natural language. It encodes watermarks at the LLM decoding stage by splitting the vocabulary into green/red lists and guides the decoding to primarily select tokens from the green list; (ii) \textbf{\textit{REMARK-LLM}}~\cite{zhang2024remark} is a neural-based watermarking scheme for LLM-generated texts. We leverage CodeT5 as the watermark insertion backbone and take both the original code and watermarking signature as input. A watermark extraction module is used to decode the message, which is trained end-to-end with the insertion module to encourage close semantics and successful message extraction.
State-of-the-art code watermarking baselines: (iii) \textbf{\textit{SWEET}}~\cite{lee2023wrote} is an inference-based watermarking scheme for LLM-generated code. It optimizes KGW's decoding by setting an entropy threshold and encodes watermarks only toward high-entropy token decoding; (iv) \textbf{\textit{SrcMarker}}~\cite{yang2024srcmarker}: is a neural-based watermarking scheme for LLM-generated code. It employs shadow transformers for watermark insertion/extraction. The watermark insertion generates both syntactic and variable rename probabilities to transform the original code into watermarked code. 

Note that we skip baselines that encode watermarks with manually engineered transformation rules like ACW~\cite{li2024resilient} and CodeMark~\cite{sun2023codemark}, which are fragile to systematic code refactoring attacks~\cite{lee2023wrote}.

\paragraph{\textbf{Syntactic Transformations for $p_{syn}$}} We follow SrcMarker's~\cite{yang2024srcmarker} syntactic transformations to construct $p_{syn}$ for C++ and Java. SrcMarker~\cite{yang2024srcmarker} did not watermark Python, and we added the following syntactic transformations: (i) Naming style: choice of naming conventions: PascalCase, camelCase, snake\_case, \_underscore\_init, or ALL\_CAPS; (ii) Operator substitution: choice between regular assignment (e.g., $i = i+1$) and augmented assignment (e.g., $i++$) for operations; (iii) Loop type: choice between using a \texttt{for} loop or a \texttt{while} loop; (iv) Nested conditions: use of merged conditions versus explicitly nested conditions; and (v) Parentheses in conditions: option to include or omit parentheses in \texttt{if} or \texttt{while} conditions.
Note that developing more syntactic transformations is orthogonal to \sys's contribution in leveraging pre-trained CodeT5 to extract better code features, and can be combined to further improve its performance.

\paragraph{\textbf{Implementation Details}}~\label{subsec:imple_detail}
Our code is implemented using PyTorch~\cite{pytorch}. The training and inference of our watermarking models are performed on NVIDIA RTX A6000 GPUs with Ubuntu 20.04.5 LTS and Intel(R) Xeon(R) Gold 6338 CPU.  
\sys{} is pre-trained on CodeSearchNet~\cite{husain2019codesearchnet}, which collects the open-source non-fork repositories from GitHub and cleans the dataset for executable functions.  \sys{} is trained for 20 epochs with the batch size set to 16, token length set to 512, $\sigma_p$ = 0.1, ($w_f$, $w_d$, $w_r$) = (1, 1, 0.05). The watermark bit length set to 4, which provides sufficient watermarking strength according to Section~\ref{subsec:exp:strength}. The neural networks are optimized by an AdamW~\cite{loshchilov2017decoupled} optimizer with the learning rate set to 5e-5. $\mathbf{R}_f$, $\mathbf{R}_m$, $\mathbf{R}_e$, $\mathbf{S}_{d1}$, and $\mathbf{S}_{d2}$ are all linear layers, which map input feature vectors of dimension 2304, 4, 768, 1536, and 1536 to outputs of dimension 768, 768, 4, 320, and 32100, respectively.
We pre-train SrcMarker~\cite{yang2024srcmarker} and REMARK-LLM~\cite{zhang2024remark} on the same dataset and report their respective performance. 
KGW~\cite{kirchenbauer2023reliability} and SWEET~\cite{lee2023wrote} encode watermarks during inference. To ensure fair comparisons, we use Qwen2.5-Coder-14B~\cite{hui2024qwen2} as the code generation model and report the conditional pass rate of watermarked code restricted to instances where the corresponding non-watermarked output is functional.

\rerevision{Following state-of-the-art code watermarking works, SWEET~\cite{lee2023wrote} and SrcMarker~\cite{yang2024srcmarker}, we provide comprehensive evaluations of watermarking performance across multiple programming languages, including Python, Java, and C++, in Sections~\ref{subsec:wm_perform}--\ref{subsec:multilang}. We evaluate \sys's robustness against variable-renaming, refactoring, and re-watermarking attacks in Section~\ref{subsec:robustness}. Furthermore, for ZKP-specific overhead analysis, we report \sys{}'s overall performance across different input samples, programming languages, watermark lengths, and decoder sizes, together with a detailed performance breakdown, in Section~\ref{subsec:zkp_perform}. Additional ablation studies and analysis are in Section~\ref{subsec:ablation}.}

\subsection{Watermark Detectability and Fidelity Performance}~\label{subsec:wm_perform}

\begin{table*}[!ht]
\centering
\resizebox{\textwidth}{!}{%
\begin{tabular}{@{}crlcccccccccccccc@{}}
\toprule
\multicolumn{2}{c}{\multirow{2}{*}{\textbf{Method}}} &
   &
  \multicolumn{4}{c}{\textbf{\textsc{HumanEval}~\cite{chen2021evaluating}}} &
  \textbf{} &
  \multicolumn{4}{c}{\textbf{\textsc{MBPP}~\cite{austin2021program}}} &
   &
  \multicolumn{4}{c}{\textbf{\textsc{EvalPlus}~\cite{evalplus}}} 
 
  \\ \cmidrule(lr){4-7} \cmidrule(lr){9-12} \cmidrule(lr){14-17}
\multicolumn{2}{c}{}                                                  &  & Pass Rate $\uparrow$ & AUROC $\uparrow$ & TPR $\uparrow$ & FPR $\downarrow$ &  & Pass Rate $\uparrow$ & AUROC $\uparrow$ & TPR $\uparrow$ & FPR $\downarrow$ &  & Pass Rate $\uparrow$ & AUROC $\uparrow$ & TPR $\uparrow$ & FPR $\downarrow$\\ \midrule

& non-WM Code & & 100\% & 0.50 & - & - & & 100\%  & 0.50 & - & - &&  100\% & 0.50 & - & -\\
\midrule
\multicolumn{17}{l}{\textbf{\textit{Natural Language}}}     \\
& KGW &  & 42.62\%& 0.82 & 0.56 & 0.03 & & 57.30\% & 0.78 & 0.43 & 0.03  & & 54.02\% & 0.73 & 0.35 & 0.05 \\
& REMARK-LLM & & \textcolor{failed}{0\%} & \textcolor{failed}{0.97} & \textcolor{failed}{0.88} & \textcolor{failed}{0.04} & & \textcolor{failed}{0\%}  & \textcolor{failed}{0.98} & \textcolor{failed}{0.89} & \textcolor{failed}{0.05}  & & \textcolor{failed}{0\%} & \textcolor{failed}{0.98} & \textcolor{failed}{0.96} &  \textcolor{failed}{0.05} \\ \midrule
\multicolumn{17}{l}{\textbf{\textit{Code}}}     \\
& SWEET &  & 82.53\% & 0.87 & 0.59 & \textbf{0.02} & & 90.00\% & 0.86 & 0.47 & \underline{0.05}  & & 84.90\% &  0.86 &  0.52 & \textbf{0.04}\\ 
& SrcMarker && \underline{95.12\%} & \underline{0.90}&  \underline{0.76} & 0.07 && \textbf{97.95\%} &  \underline{0.91} &   \underline{0.76} & 0.06 && \textbf{95.57\%} & \underline{0.92} & \underline{0.81} & 0.07  \\
& \sys{} && \textbf{95.12\%} & \textbf{0.97} & \textbf{0.98} & \underline{0.06} && \underline{97.64\%} & \textbf{0.97} & \textbf{0.99} & \textbf{0.05} && \underline{95.39\%} & \textbf{0.97} & \textbf{0.98} & \underline{0.06}  \\\bottomrule

\end{tabular}%

}
\caption{ \sys{} performance on watermarking HumanEval~\cite{chen2021evaluating}, MBPP~\cite{austin2021program}, and EvalPlus~\cite{evalplus} datasets when comparing with natural language watermarking KGW~\cite{kirchenbauer2023watermark} and REMARK-LLM~\cite{zhang2024remark}; code watermarking SWEET~\cite{lee2023wrote} and \rerevision{SrcMarker~\cite{yang2024srcmarker}}.  The best metric values are highlighted in \textbf{bold} text, the second best metric values are \underline{underlined}, and
\textcolor{failed}{grey} means failed watermark insertion (0\% pass rate).}
\label{tab:table_main}
\vspace{-15pt}
\end{table*}


\subsubsection{Comparison with Prior Work} The \sys's watermarking performance is in Table~\ref{tab:table_main}.
We highlight that \sys{} is able to provide high detectability while maintaining code functionality invariant.  

\textbf{Compared to KGW~\cite{kirchenbauer2023reliability}} KGW watermarks LLM-generated code by promoting token decoding from the green list of the vocabulary. We relax the green/red list split ratio $\gamma$ to 0.5 and set the constant $\delta$ added on green tokens' probabilities to 3 to guide watermark insertion on green lists while ensuring Code LLM generates compilable code. As syntactic tokens (e.g., \texttt{for} and \texttt{while}) are sensitive to alterations, restricting the watermark insertion at the token-level results in an average of 48.69\% pass rate drop to achieve an average of 0.78 AUROC for watermark detection.

\textbf{Compared to REMARK-LLM~\cite{zhang2024remark}} REMARK-LLM  primarily replaces words with their synonyms or changes the sentence syntax for watermark insertion. It was able to learn code semantics by leveraging CodeT5~\cite{wang2021codet5} as the watermark insertion backbone and fine-tuning on code datasets. However, it transforms code without syntactic constraints, making the watermarked code non-compileable. As such, while reaching significant detectability, REMARK-LLM has low pass rates that corrupted the watermarked codes' functionality.

\textbf{Compared to SWEET~\cite{lee2023wrote}} SWEET improves over KGW by restricting watermark decoding on high-entropy tokens for better detectability and fidelity. However, encoding watermarks at the token level loses the global view of the code and results in potential syntactic invalidity. As such, it demonstrates a 10.24\% drop in pass rate and an average of 10.99\% lower AUROC for watermark detection compared to \sys{}.

\textbf{Compared to \rerevision{SrcMarker~\cite{yang2024srcmarker}}} SrcMarker uses shallow transformers trained from scratch for watermark insertion, which limits its code feature extraction ability.  
\sys{}, on the other hand, leverages CodeT5~\cite{wang2021codet5} as the backbone to extract better code features for improved watermark insertion performance, which results in an average of 6.59\% higher AUROC scores and 26.61\% higher TPR among all benchmarks than SrcMarker. Besides, both \sys{} and SrcMarker 
insert watermarks by applying syntactic transformations and variable renaming that preserve compilation correctness, resulting in less than a 5\% drop in pass rate.

\subsubsection{Watermarking Strength}~\label{subsec:exp:strength}  
To quantify statistical evidence against the null hypothesis of no watermark presence, we compute a one-sided $z$-score from the aggregated bit matches across the dataset following Equation~\ref{eq:zscore}. Using a one-tailed significance level of 0.05, the rejection threshold is  $z\ge 1.645$. 
When $F$ is set to 1, the average $z$ is 1.92, which already exceeds this threshold. Therefore, it provides statistically significant evidence of watermark presence and demonstrates that \sys{} yields sufficient watermarking strength to prove ownership. 

The watermark would be further strengthened when multiple functions are stacked, which is a more realistic setting when a feature requires multiple code functions to implement. Say we watermark a file with $F=5$ and \sys{} encode 4-bit signatures on every function. Then, the z-score would be boosted to 4.29 with the one-tailed significance level to be $1.8\times10^{-5}$, providing much stronger support for ownership verification. 

\subsection{Multi-lingual Code Watermarking}~\label{subsec:multilang}
In Table~\ref{tab:languages}, we compare \sys's ability to watermark other coding languages (C++ and Java) with the best-performing baseline SrcMarker. The pre-trained CodeT5 model effectively extracts code features from different coding languages for high-quality watermark insertion. As such, \sys{} consistently achieves high detectability and functionality-preservation after watermarking. \sys{} averagely improves AUROC by 3.78\% and TPR by 18.90\% over SrcMarker~\cite{yang2024srcmarker}, while maintaining a comparable FPR. Notably, both methods achieve over 98\% pass rate, demonstrating that watermarking introduces minimal disruption to code functionality.

\begin{table}[!ht]
    \centering
    \begin{tabular}{c|cc|cc}
    \toprule
    \multirow{2}{*}{Metrics} &  \multicolumn{2}{c|}{MBCPP (C++)} & \multicolumn{2}{c}{MBJP (Java)}\\\cline{2-5}
      & SrcMarker & \sys{} & SrcMarker & \sys{} \\\hline
Pass Rate $\uparrow$ & 98.69\% & \textbf{98.95\%}  & 98.69\% & \textbf{98.69\%}\\
AUROC $\uparrow$ & 0.93 & \textbf{0.96} & 0.92 & \textbf{0.96}\\
TPR $\uparrow$ & 0.80 & \textbf{0.98} & 0.84 &  \textbf{0.97}\\
FPR $\downarrow$ & \textbf{0.06} & 0.07 & \textbf{0.06} & 0.07\\
    \bottomrule     
    \end{tabular}
    \caption{Multi-lingual code watermarking performance.}
    \label{tab:languages}
\end{table}


\subsection{Zero Knowledge Watermark Verification Overhead}~\label{subsec:zkp_perform}

\subsubsection{\revision{Overall Performance}} We benchmark our ZK watermark extraction and verification by describing algorithm \ref{alg:extract} in a computational graph that can be easily translated into a zero-knowledge circuit. The computation is done using the EZKL framework \cite{ezkl} in conjunction with the Halo2 proof system \cite{halo2_repo}, resulting in a small zk-SNARK proof that any third-party arbitrator can easily verify. Our approach ensures that no information is leaked about the parameters of the decoder $\bf{R}_e$ and the watermarking methodology, including the original watermark $M$. The verification overhead is in Table~\ref{tab:zkp}. Generating the zk-SNARK proof only takes the prover \Prv \rerevision{\textbf{7.154 seconds}}, while only requiring approximately \textbf{2.6 GB} of RAM. While this is significantly slower than standard inference, we highlight that this is a fully privacy-preserving solution, and, more importantly, proof generation only has to be done once. This process results in a proof $\pi$ of size \rerevision{\textbf{17.12 KB}}, which can be transmitted to as many verifiers as necessary. We benchmarked the verification over 25 runs. This proof can be verified by any verifier \Vrf in an average of \rerevision{\textbf{118.6 milliseconds}}, while only requiring a maximum of approximately \rerevision{\textbf{222.5 MB}} of RAM. All the verifier needs to verify a proof is the proof $\pi$ and the verifier key $\mathcal{VK}$, which is only \textbf{577.3 KB} in our setup. This results in a required communication cost of less than a megabyte. As the proof generation time is amortized due to it only being performed once, \sys{}'s watermark extraction and verification scheme is an extremely communication and runtime-efficient solution that ensures the security of watermarks that are applied to LLM-generated code.


\rerevision{
\begin{table}[!ht]
    \centering
    \begin{tabular}{c | c c c c}
         \toprule
        & Frequency & Comm. Size & RAM & Time \\
    \midrule
        Proof Gen. & Once & Proof Size: 17.12\,KB & 2.618 $\pm$ 0.053\,GB & 7.154 $\pm$ 0.438\,s \\
        WM Verif. & Every WM & Vrf Key Size: 577.3\,KB & 222.50 $\pm$ 0.1\,MB & 118.6 $\pm$ 4.6\,ms \\
    \bottomrule
    \end{tabular}
    \caption{Zero-knowledge proof generation (Proof Gen.) and watermark verification (WM Verif.) overhead of the deployed circuit, reported as mean $\pm$ std across 25 runs.}
    \label{tab:zkp}
\end{table}}

\subsubsection{\revision{Performance Breakdown and Analysis}} \revision{We analyze how each optimization component in Section~\ref{subsec:method:verification} contributes to \sys's zero-knowledge verification performance in Table~\ref{tab:verify_breakdown}. The baseline computes the bit error rate on a revealed $M^\prime$, which is cheap to verify the watermark within \rerevision{\textbf{118.2\,ms}} but unacceptable for deployment. \rerevision{Optimizations \circled{1} and \circled{2} leave the verification cost statistically unchanged --- the three off-chain-BER rows agree within one standard deviation, and their proof sizes differ by less than 0.1 KB --- since at the deployed decoder scale the circuit cost is dominated by thefixed logrows budget rather than the activation or parameter precision. Instead, their benefit is structural: the polynomial ReLU \circled{1} arithmetizes to pure multiply-and-add gates and eliminates the range check lookup table, which is what later permits fusing zkFeedForward and zkBER into a single computational graph in \circled{4}, while Bfloat16 quantization \circled{2} halves parameter storage on the proving side without affecting watermark extraction accuracy.} Optimization \circled{3} enhances the security of $M^\prime$ by including bit error rate computation into the zero-knowledge proof with a custom Halo2 zkBER circuit. It binds $M^\prime$ to $\{0,1\}^n$ in ZKP to close the forgery vector of the baseline. Realized as a separate circuit, however, it requires verifying two proofs and roughly doubles verification time to \rerevision{\textbf{235.0\,ms}}. To mitigate the overhead while ensuring the watermark privacy, Optimization \circled{4}, the composability layer, fuses \texttt{zkFeedForward} and \texttt{zkBER} into a single computational graph, which reclaims \rerevision{\textbf{$-116.4$\,ms}} and returns verification to a single proof. The deployed configuration thus verifies in \rerevision{\rerevision{\textbf{118.6\,ms}}} ---  \rerevision{within 0.4 ms of the insecure baseline and one combined standard deviation of its mean, i.e. statistically indistinguishable} --- while being the only configuration that simultaneously keeps $M^\prime$ private, binds it to a binary signature, and verifies a single proof. The composability layer contributes to the largest performance gain, converting the otherwise undeployable two-proof secure design into one whose verification cost matches the naive baseline. This demonstrates that through system-level co-optimization, \sys{} preserves zero-knowledge verification efficiency while securing the confidentiality and unforgeability of the underlying watermarks.}

\rerevision{
\begin{table}[!ht]
    \centering
    \small
    \begin{tabular}{l | c c c c}
        \toprule
            Variant & Verify time & Verify RAM & Proof KB \\
        \midrule
            Baseline (ReLU, fp32; BER off-chain, reveals $M'$) & 118.2 $\pm$ 3.9\,ms & 222.8 $\pm$ 1.6\,MB & 17.48 \\
            + (1) polynomial-approx ReLU  $x^2+x$ & 123.6 $\pm$ 8.3\,ms & 224.4 $\pm$ 3.4\,MB & 17.50 \\
            + (2) Bfloat16 parameter quantization & 118.4 $\pm$ 3.7\,ms & 222.2 $\pm$ 3.5\,MB & 17.55 \\
            + (3) custom Halo2 zkBER (binds $M'$ in-ZK) & 235.0 $\pm$ 6.7\,ms & 223.5 $\pm$ 2.6\,MB & 34.44 \\
            + (4) composability layer -- deployed$^\dagger$ & 118.6 $\pm$ 4.6\,ms & 222.5 $\pm$ 0.1\,MB & 17.12 \\
        \bottomrule
    \end{tabular}
    \caption{WM-verification cost broken down by RoSeMary's four section \ref{subsec:method:verification} ZK co-optimizations, layered cumulatively on a strawman baseline ($R_e$ in ZK, BER computed off-chain on the revealed $M^\prime$ -- cheapest verify, but leaks $M^\prime$ and does not bind the prover to a binary signature).  (1)~polynomial-approx ReLU $\sigma(x)=x^2+x$ replaces the ReLU lookup with an arithmetic gate; (2)~Bfloat16 parameter quantization round-trips weights through bf16 on the proving side; (3)~custom bitwise-AND zkBER as its own Halo2 sidecar moves BER into ZK and binds $M^\prime$ to $\{0,1\}^n$, paying a two-proof verify-time tax; (4)~composability fuses zkFeedForward and zkBER back into one graph -- the deployed configuration is the only row simultaneously providing in-ZK BER, binary-signature soundness, and single-proof verification.  Verify time is the ezkl-internal cryptographic-verify cost; Verify RAM is the peak RSS of `ezkl verify'; both reported as mean $\pm$ std across 25 per-variant trials. The $\dagger$ denotes the deployed configuration as seen in Table \ref{tab:zkp}}
\label{tab:verify_breakdown}
\end{table}
}


\revision{Additionally, we analyze how different decoder sizes and signature lengths, and programming languages would impact the overall proof performance. As seen in Table~\ref{tab:zkp_sweep}, the three axes have markedly different effects: the decoder hidden width is the dominant cost driver, the watermark length is essentially free, and the verification cost is language invariant by design.
}
\revision{
Increasing the decoder hidden width from 256 to 512 yields only a modest increase in cost, but the 1024 wide decoder crosses a power of two domain boundary in the underlying Halo2 circuit -- the $\mathrm{logrows}$ budget rises from 18 to 19 -- which doubles both the proof size, and the verification key, and inflates the verification time to \rerevision{156.1\,ms}. This justifies \sys's choice in Section~\ref{subsec:method:verification} to quantize $\mathbf{R}_e$ to a compact shallow decoder: a deeper or wider decoder would push the verifier across this boundary and roughly double the artifact sizes that must be transmitted and loaded per verification.
}
\revision{
Varying the watermark length from 4 bits to 16 bits leaves verification time statistically unchanged \rerevision{(118.6~ms at 4,8, and 16 bits, all within $1\sigma$ of each other)}, the verification key identical, and the proof size essentially unchanged \rerevision{(17.08 to 17.12~KB)}. Cryptographic verification cost therefore does not trade off against watermark strength: an owner can lengthen the signature to drive down false positive rates and tighten the $z$-score of Equation~\ref{eq:zscore} at no measurable ZK expense.
}
\revision{
Finally, the three programming languages (Python, Java, C\texttt{++}) produce verification numbers that are indistinguishable within noise. This is by design rather than by coincidence: per Section~\ref{subsec:method:verification}, the language specific encoder $\mathbf{S}_e$ is evaluated off-chain and only the fixed width decoder $\mathbf{R}_e$ enters the circuit which consumes a 768 dimensional feature vector regardless of source language. Consequently \sys{} verifies watermarks across programming languages at \emph{identical} cryptographic cost, in contrast to the watermark insertion stage which is necessarily language dependent.
}
\rerevision{
\begin{table}[!ht]
    \centering
    \small
        \begin{tabular}{l c c c | c c c c}
        \toprule
            Axis & feat.\ dim & hidden & $n_\text{bits}$ & Proof time & Verify time & Proof KB & VK KB \\
        \midrule
            decoder size & 768 & 256 & 4 & 5.95\,s & 116.7 $\pm$ 4.7\,ms & 17.09 & 577.3 \\
            decoder size$^\dagger$ & 768 & 512 & 4 & 7.15\,s & 118.6 $\pm$ 4.6\,ms & 17.12 & 577.3 \\
            decoder size & 768 & 1024 & 4 & 11.96\,s & 156.1 $\pm$ 9.2\,ms & 29.31 & 1154.2 \\
            \hline
            watermark length$^\dagger$ & 768 & 512 & 4 & 7.15\,s & 118.6 $\pm$ 4.6\,ms & 17.12 & 577.3 \\
            watermark length & 768 & 512 & 8 & 6.75\,s & 118.6 $\pm$ 3.5\,ms & 17.08 & 577.3 \\
            watermark length & 768 & 512 & 16 & 7.38\,s & 118.6 $\pm$ 3.4\,ms & 17.10 & 577.3 \\
            \hline
            language (python)$^\dagger$ & 768 & 512 & 4 & 7.15\,s & 118.6 $\pm$ 4.6\,ms & 17.12 & 577.3 \\
            language (java) & 768 & 512 & 4 & 7.13\,s & 116.9 $\pm$ 4.6\,ms & 17.10 & 577.3 \\
            language (c\texttt{++}) & 768 & 512 & 4 & 6.70\,s & 128.6 $\pm$ 10.1\,ms & 17.06 & 577.3 \\
        \bottomrule
    \end{tabular}
    \caption{ZKP cost vs.\ decoder hidden size, watermark length, and source language.  Every cell is the deployed section-4.3 circuit (poly-ReLU + Bfloat16 + fused zkFeedForward/zkBER). Rows marked $\dagger$ are the deployed configuration as seen in Table \ref{tab:zkp}.}
    \label{tab:zkp_sweep}
\end{table}
}

\subsection{Robustness Evaluations}~\label{subsec:robustness}

As in Section~\ref{sec:threat}, we assume the adversary is an end-user who adopts LLM-generated code for malicious purposes. Their goal is to preserve the code’s functionality while removing the embedded signatures to evade detection. The adversary has access only to the watermarked code and cannot manipulate the LLM owner’s code generation or watermarking process. They may discover the signatures if verification is performed without a ZKP circuit, but when the signature is provided as a private input within the ZKP circuit, it remains hidden from the adversary. \revision{Consistent with the claim taxonomy in Section \ref{zkp_guarantees}, the results in this section are empirical robustness measurements under the stated attacks, and are distinct from the cryptographic guarantees of the verification protocol.}
As a proof of concept, we evaluate attack performance on the MBPP dataset~\cite{austin2021program}. 

\subsubsection{Watermark Removal Attack}~\label{exp:subsubsec:wm_removal}
Following Section~\ref{subsec:problem:threat}, we consider:
(i) \textbf{Variable-Rename Attack} (VA): the adversary randomly renames variables using semantically similar words to preserve both functionality and readability. We modify 10\%, 30\%, and 50\% of the variables in each function. (ii) \textbf{Refactor Attack} (RA): the adversary refactors the watermarked code using open-source code LLMs, which is able to provide comparable accuracy but more code refactoring variants compared to traditional code refactoring tools~\cite{oueslati2025refagent}.
Here, we instruct Qwen2.5-Coder-32B-Instruct~\cite{hui2024qwen2} model to refactor the code using the prompt: ``You are Qwen Coder. You are a helpful assistant that performs code refactoring to remove watermark. Please only return the refactored code.'' RA is a compositional transformation that can simultaneously apply multiple semantics-preserving edits (e.g., reformatting, control-flow restructuring, try/catch wrapping, and variable renaming) while attempting to keep the original behavior unchanged. We use RA as a practical proxy for real-world post-processing pipelines that developers may apply before reuse of the watermarked code.
(iii) \revision{\textbf{Re-Watermarking Attacks} (RWA): The adversary may have prior knowledge of the watermarking system, and trains a surrogate \sys{} with a different set of hyperparameters. With the trained surrogate watermarking system, the adversary aims to encode his/her new watermarks and deactivates the ownership proof of the original code LLM.}

\begin{figure}[!ht]
    \centering
    \includegraphics[width=0.7\columnwidth]{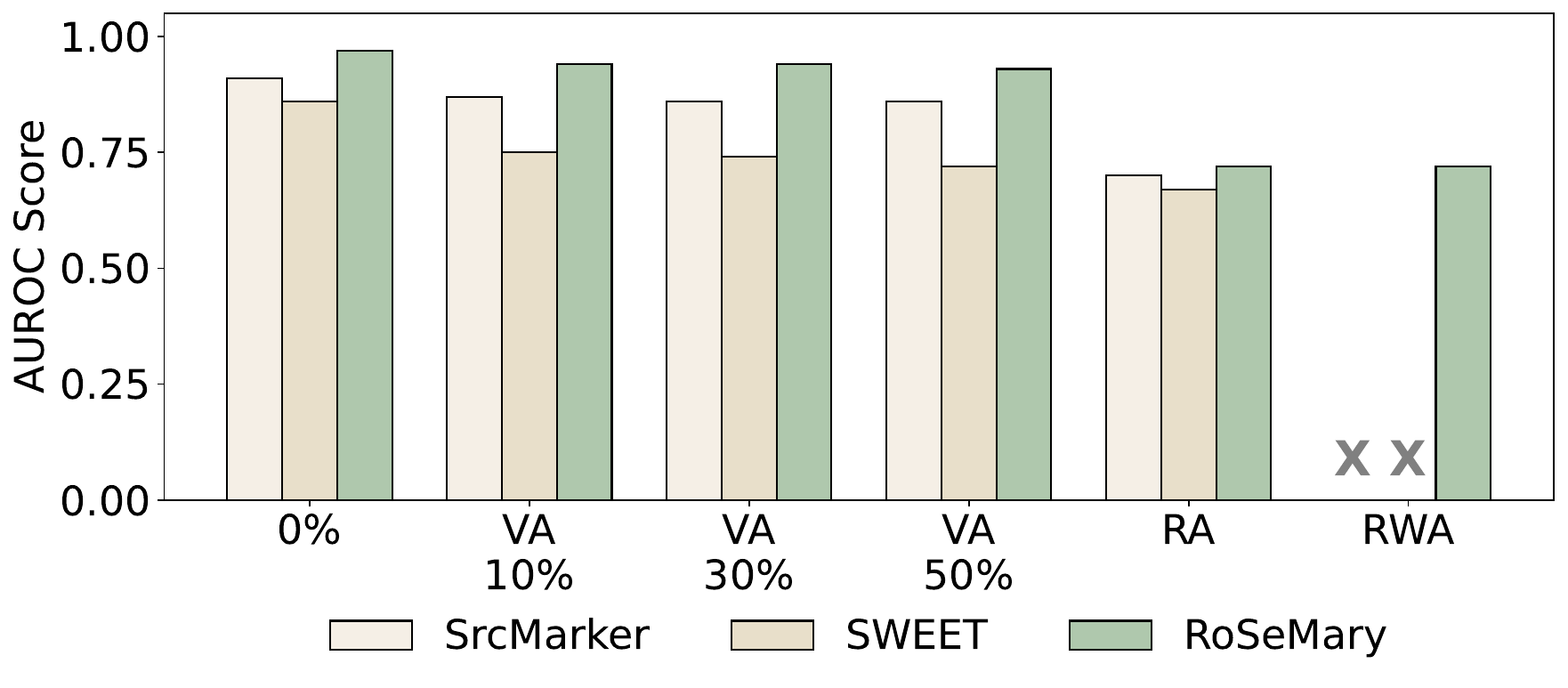}
    \caption{\revision{Robustness evaluation results under different attacks.}}
    \label{fig:attack}
\end{figure}

The attack results are shown in Figure~\ref{fig:attack}. We compare \sys{} with \rerevision{SrcMarker~\cite{yang2024srcmarker}} and SWEET~\cite{lee2023wrote}. \sys{} maintains an AUROC above 0.93 under variable-rename attacks even when 50\% of variable names are replaced, whereas SrcMarker and SWEET show AUROC decreases of 0.07 and 0.21, respectively. Under refactor attacks, \sys{} achieves an AUROC of 0.73 when being refactored by Qwen2.5-Coder and demonstrates strong robustness. This robustness arises mainly from two factors: (i) leveraging CodeT5~\cite{wang2021codet5} for improved feature extraction to enhance detectability, and (ii) adversarial training, which enables the extraction module to learn malicious transformations and accurately recover watermark signatures. 

\revision{For re-watermarking attacks, we train a surrogate model that has the same end-to-end watermarking pipeline as \sys{}, but change the hyperparameters to enforce different transformation behaviors. Since RWA is an adaptive attack that targets only \sys's pipeline, we only show \sys's performance in Figure~\ref{fig:attack}. As seen, \sys{} is able to maintain 0.72 AUROC under attacks, which demonstrates its resilience even when the adversary has prior knowledge of the system. }

\subsubsection{\revision{Attack Performance on Different Programming Languages}} \revision{Additionally, we evaluate the performance of Java and C++ programs watermarked by \sys{} under refactoring attacks. Following the setup in Section~\ref{exp:subsubsec:wm_removal}, we use DeepSeek-Coder-7B~\cite{deepseek-coder} to refactor the watermarked code and report the results in Table~\ref{tab:multi_lingual_attack}. As shown, \sys{} achieves a similar level of watermark detectability on Java and C++ programs, \rerevision{where it has a TPR of 0.66 on C++ and 0.54 on Java under 3/4 matching-bit threshold}. It demonstrates \sys's generalizability and robustness in providing strong ownership proof for code beyond Python. }

\begin{table}[!ht]
    \centering
    \begin{tabular}{c|cc|cc}
    \toprule
    \multirow{2}{*}{Metrics}
    & \multicolumn{2}{c|}{MBCPP (C++)}
    & \multicolumn{2}{c}{MBJP (Java)} \\
    \cline{2-5}
    & Original & Attack & Original & Attack \\
    \hline
    AUROC $\uparrow$
    & 0.96 & 0.72
    & 0.96 & 0.64 \\

    \rerevision{TPR $\uparrow$}
    & \rerevision{0.98}
    & \rerevision{0.66}
    & \rerevision{0.97}
    & \rerevision{0.54} \\

    \rerevision{FPR $\downarrow$}
    & \rerevision{0.07}
    & \rerevision{0.30}
    & \rerevision{0.07}
    & \rerevision{0.30} \\
    \bottomrule
    \end{tabular}
    \caption{\revision{Multi-lingual code watermarking performance under
    refactoring attacks.}
    \rerevision{For \texttt{Attack}, TPR and FPR are reported using a 3/4 matching-bit threshold.}
    }
    \label{tab:multi_lingual_attack}
\end{table}

\subsubsection{Watermark Forgery Attack}
\sys{} is also resilient to \textbf{forgery attacks}, where an adversary attempts to produce a proof that validates an ownership claim without possessing the owner’s secret signature. In our threat model, end users only have access to the released watermarked code, while the decoding module, seed, and signature remain confidential to the owner. Moreover, verification is performed via zero-knowledge proofs, where the signature is provided as a private input to the circuit and is not disclosed to the arbitrator or any third party. As a result, an adversary cannot learn the signature from the verification, nor can a malicious arbitrator exploit the verification process to impersonate the owner and generate false ownership claims.

\subsection{Ablation Study and Analysis}~\label{subsec:ablation}

\subsubsection{Watermark Insertion Overhead} The time taken for watermark insertion is in Table~\ref{tab:overhead}, which is the average overhead for encoding signatures onto 50 examples from the MBPP~\cite{austin2021program} dataset. \sys{} embeds watermarks 90\% faster than SWEET's inference-based approach, which requires entropy calculation for every token and splits the vocabulary into green/red lists for high-entropy tokens. Compared to SrcMarker, while \sys{} introduces more complex architectures for better watermarking performance, the additional watermark insertion overhead is less than 0.01s per sample. As such, \sys's watermark insertion is efficient. 

\begin{table}[!ht]
    \centering
    \begin{tabular}{c|ccc}
    \toprule
       Method  & SWEET & SrcMarker & \sys{} \\\hline
       Time (s)  & 0.234 & 0.021 & 0.027 \\
    \bottomrule
    \end{tabular}
    \caption{Different frameworks' watermark insertion overhead.}

    \label{tab:overhead}
\end{table}

\subsubsection{Impact of Training Loss Weights}~\label{subsec:loss_impact} The impact of different loss weight choices on watermark performance is in Table~\ref{tab:loss_weight}. We train \sys{} on the CodeSearchNet~\cite{husain2019codesearchnet} and evaluate its performance on MBPP~\cite{austin2021program}. Weighting more on the detectability loss $w_d$ leads to better detection scores (AUROC, TPR, and FPR) than the robustness loss ($w_r$), as excessive focus on robustness against malicious modifications may hinder extraction performance on unaltered code. Additionally, increasing the weight of the functionality loss ($w_f$) maintains a comparable pass rate to the other configurations.

\begin{table}[h!]
    \centering
    
    \begin{tabular}{c|cccc}
    \toprule
     ($w_f$, $w_d$, $w_r$) &  Pass Rate $\uparrow$ & AUROC $\uparrow$ & TPR $\uparrow$ & FPR $\downarrow$ \\ \hline
     (0.8, 0.1, 0.1)    &  97.74\% & 0.95 & 0.93 & 0.07\\
     (0.1, 0.8, 0.1)    & \textbf{97.74\%} & \textbf{0.97} & \textbf{0.98} & \textbf{0.05}\\
     (0.1, 0.1, 0.8)    &  97.74\% & 0.97 & 0.95 & 0.05\\
    
    \bottomrule
    \end{tabular}
    \caption{Impact of loss weights on \sys's performance.}
    \label{tab:loss_weight}
\end{table}

\subsubsection{Impact of Decoder $\textbf{S}_{d1}$ and $\textbf{S}_{d2}$}~\label{subsec:decoder_impact} We analyze the impact of training with multiple decoders in Table~\ref{tab:decoder}. We pre-train \sys{} on the CodeSearchNet~\cite{husain2019codesearchnet}, with either $\textbf{S}_{d1}$  or $\textbf{S}_{d2}$, and evaluate the performance on MBPP~\cite{austin2021program}. As seen, encoding watermarks with sole syntactic transformations ($\textbf{S}_{d1}$) or variable name transformations ($\textbf{S}_{d2}$) results in close pass rates but degraded detectability, as less information can be embedded onto the code for watermark insertion. 
Besides, encoding only syntactic transformations results in higher detectability. Because such transformations carry more information for the signature feature to encode into the watermarked code. 

\begin{table}[h!]
    \centering
    \begin{tabular}{cc|cccc}
    \toprule
     $\textbf{S}_{d1}$ & $\textbf{S}_{d2}$ &  Pass Rate $\uparrow$ & AUROC $\uparrow$ & TPR $\uparrow$ & FPR $\downarrow$  \\ \hline
     \cmark & \xmark &  98.05\% &  0.81 &  0.36 & 0.06 \\
     \xmark & \cmark & \textbf{98.77\%}  & 0.73  & 0.24 &  0.06   \\
     \cmark & \cmark &  97.64\% & \textbf{0.97} & \textbf{0.99} &  \textbf{0.05}\\
    \bottomrule
    \end{tabular}
    \caption{Impact of decoders on \sys's performance.}
    \label{tab:decoder}
\end{table}

\subsubsection{Watermark Examples}~\label{subsec:wm_example} We show the watermark examples in Figure~\ref{tab:example}, where Listing~\ref{lst:original},\ref{lst:original2} are original code and Listing~\ref{lst:wm},\ref{lst:wm2}  are watermarked one. \sys{} will transform the variable names and syntactic structure with meaningful content while maintaining the functionality invariant. For example, Listing~\ref{lst:original}'s two if statements (line 6 and line 7) are merged into one (line 6) in Listing~\ref{lst:wm}. 
The function name is updated from \texttt{get\_closest\_vowel} to \texttt{getClosestVowel}.   In Listing~\ref{lst:wm2}, the variable name \texttt{i} is changed to \texttt{item}, and parentheses are added for \texttt{(t>0)} in line 8.

\begin{figure*}[!ht]
    \centering
    \begin{minipage}[t]{\textwidth}
    \begin{minipage}[t]{0.25\textwidth}
\centering
 \begin{lstlisting}[style=pythonstyle,label=lst:original,caption=non-WM code \#1]
 def get_closest_vowel(word):
    if len(word) < 3:
        return ""
    vowels = {"a", "e", "i", "o", "u", "A", "E", 'O', 'U', 'I'} 
    for i in range(len(word)-2, 0, -1): 
        if word[i] in vowels: 
            if (word[i+1] not in vowels) and (word[i-1] not in vowels): 
                return word[i]
    return ""
\end{lstlisting}
\end{minipage} 
    \begin{minipage}[t]{0.25\textwidth}
        \centering
        \begin{lstlisting}[style=pythonstyle,label=lst:wm,caption=WM code \#1]
def (*@\diffadd{getClosestVowel}@*)((*@\diffadd{dates}@*)):
    if (len((*@\diffadd{dates}@*)) < 3):
        return ""
    vowels = {"a", "e", "i", "o", "u", "A", "E", 'O', 'U', 'I'}
    for i in range(len((*@\diffadd{dates}@*))-2, 0, -1):
        if ((*@\diffadd{dates[i] in vowels}@*) and (dates[i+1] not in vowels) and (dates[i-1] not in vowels)):
            return (*@\diffadd{dates}@*)[i]
    return ""
        \end{lstlisting}
    \end{minipage} 
    \begin{minipage}[t]{0.24\textwidth}
 \begin{lstlisting}[style=pythonstyle,label=lst:original2,caption=non-WM code \#2]
def histogram(test):
    dict1={} 
    list1=test.split(" ") 
    t=0 
    for i in list1: 
        if(list1.count(i)>t) and i!='': 
            t=list1.count(i) 
        if t>0: 
            for i in list1: 
                if(list1.count(i)==t): 
                    dict1[i]=t 
            return dict1
\end{lstlisting}
    \end{minipage}
    \begin{minipage}[t]{0.24\textwidth}
        \begin{lstlisting}[style=pythonstyle,label=lst:wm2,numbers=none,caption=WM code \#2]
def histogram(test):
    dict1={}
    list1=test.split(" ")
    t=0
    for (*@\diffadd{item}@*) in list1:
        if ((list1.count((*@\diffadd{item}@*))>t) and (*@\diffadd{item} @*)!=''):
            t=list1.count((*@\diffadd{item}@*))
        if (*@\diffadd{(t>0)}@*):
            for (*@\diffadd{item}@*) in list1:
                if(list1.count((*@\diffadd{item}@*))==t):
                    dict1[(*@\diffadd{item}@*)]=t
        return dict1
        \end{lstlisting}
    \end{minipage} 
    
    \end{minipage}\\
    \caption{Watermarked examples randomly selected from \rerevision{HumanEval~\cite{chen2021evaluating}} and MBPP~\cite{austin2021program} dataset. The Listing~\ref{lst:original},\ref{lst:original2} are the original code (non-WM code), and the Listing~\ref{lst:wm},\ref{lst:wm2} show the watermarked one (WM code), where all watermarks are successfully extracted.}
    \label{tab:example}
\end{figure*}

\subsection{Overall Evaluation}
\begin{figure}[t]
    \centering
    \includegraphics[width=0.4\columnwidth]
    {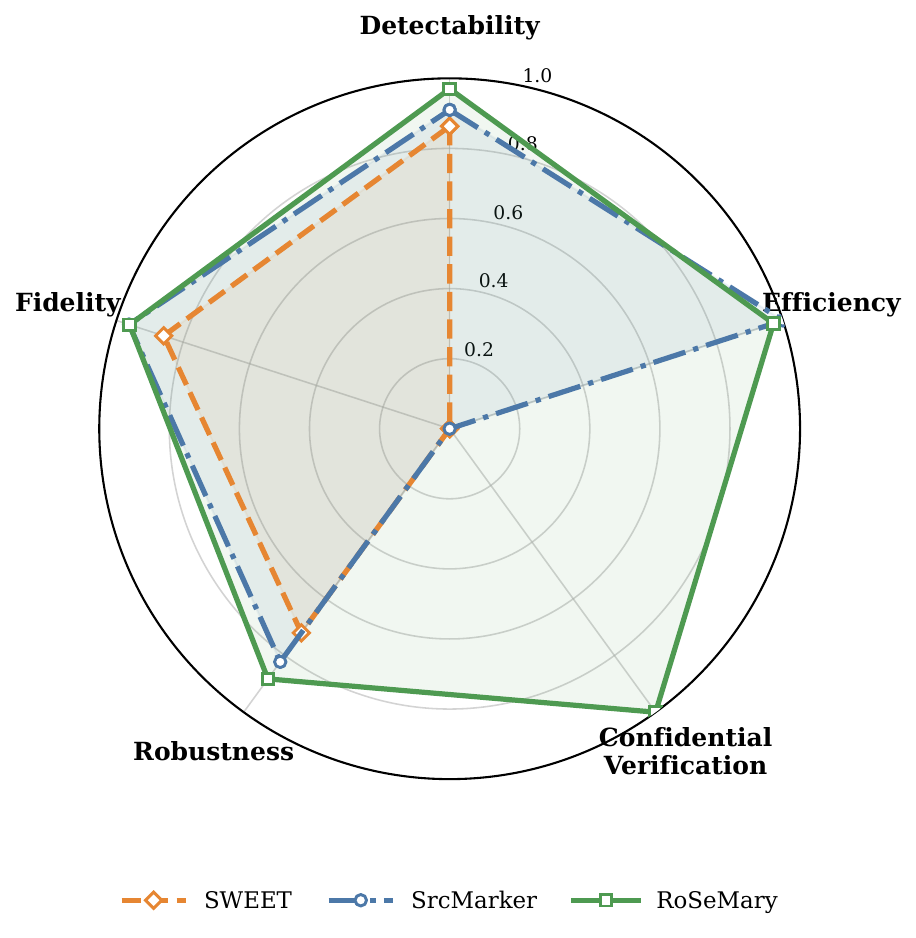}
    \caption{ \rerevision{
    Overall evaluation of \sys{}, SrcMarker, and SWEET across detectability, fidelity, robustness, watermark-insertion efficiency, and zero-knowledge
    verification. Higher values indicate better performance on every axis. \sys{} provides the most balanced coverage of the requirements for practical code-watermarking systems.
    }
    }
    \label{fig:overall_watermark_comparison}
\end{figure}

\rerevision{
Figure~\ref{fig:overall_watermark_comparison} provides a comprehensive comparison
of \sys{}, SrcMarker, and SWEET across five dimensions: detectability, fidelity,
robustness, watermark-insertion efficiency, and security with confidential verification.
Detectability and fidelity are computed as the mean AUROC and pass rate,
respectively, across the benchmarks reported in
Table~\ref{tab:table_main}. Robustness is measured as the mean AUROC under the
variable-renaming and refactoring attacks reported in
Figure~\ref{fig:attack}. Efficiency is obtained by normalizing the watermark
insertion times reported in Table~\ref{tab:overhead}, such that higher values
indicate lower insertion overhead. Zero-knowledge-based verification is
represented using a binary indicator, where $1$ denotes support for ownership
verification without revealing the embedded signature, whereas $0$ indicates
that this capability is not provided.}

\rerevision{ As shown in Figure~\ref{fig:overall_watermark_comparison}, \sys{} achieves the highest detectability and robustness while preserving fidelity comparable to
SrcMarker and markedly higher than SWEET. SrcMarker attains the lowest watermark-insertion time, although \sys{} remains close in normalized
efficiency. More importantly, \sys{} is the only evaluated framework that
supports confidential ownership verification through zero-knowledge proofs.
Overall, these results demonstrate that \sys{} provides the most balanced coverage of the requirements for practical code-watermarking systems.
}

\section{Conclusion}
In this paper, we present \sys{}, the ML/Crypto codesign framework to secure high-quality code LLM watermark verification. It trains the watermark insertion and extraction modules end-to-end, aiming to ensure the watermarked codes' functionality-invariant, while maintaining the detectability of the watermark in the adversarial environment. To guarantee signature confidentiality and preserve code usability, \sys{} customizes and co-optimizes non-interactive zero-knowledge proofs for secure watermark verification, which allows third-party legal arbitrators to verify ownership without revealing the signature to them. Extensive evaluations across various coding benchmarks demonstrate that \sys{} effectively preserves code fidelity, achieves high detectability and robustness, and supports efficient zero-knowledge verification.

\bibliographystyle{ACM-Reference-Format}
\bibliography{example_paper,references}

\end{document}